\documentclass[runningheads]{llncs}
\usepackage{amsfonts}
\usepackage{amsmath}
\usepackage{booktabs} 
\usepackage{graphicx}
\usepackage{pgf,tikz,hyperref}
\usepackage[all]{nowidow}
\usepackage[utf8]{inputenc}
\graphicspath{{./figures/}}
\DeclareGraphicsExtensions{.pdf,.png}

\definecolor{lime}{HTML}{A6CE39}
\DeclareRobustCommand{\orcidicon}{%
\begin{tikzpicture}
\draw[lime, fill=lime] (0,0)
circle [radius=0.16]
node[white] {{\fontfamily{qag}\selectfont \tiny ID}};
\draw[white, fill=white] (-0.0625,0.095)
circle [radius=0.007];
\end{tikzpicture}
\hspace{-2mm}
}

\foreach \x in {A, ..., Z}{%
\expandafter\xdef\csname orcid\x\endcsname{\noexpand\href{https://orcid.org/\csname orcidauthor\x\endcsname}{\noexpand\orcidicon}}
}

\begin{document}
\title{Being Patient and Persistent}
\subtitle{Optimizing An Early Stopping Strategy for Deep Learning in Profiled Attacks}

%
%
\author{Servio Paguada\inst{1,2}\orcidA{} \and
Lejla Batina\inst{1}\orcidB{} \and
Ileana Buhan\inst{1} \and 
Igor Armendariz\inst{1}\orcidD{}}

\authorrunning{Paguada \textit{et al.}}
%
\institute{Radboud University, Nijmegen, The Netherlands
\email{\{servio.paguadaisaula,lejla.batina,ileana.buhan\}@ru.nl}\\ \and
IKERLAN Research Center, Mondragon/Arrasate, Spain\\
\email{\{slpaguada,iarmendariz\}@ikerlan.es}}

\maketitle              
\begin{abstract}
The absence of an algorithm that effectively monitors deep learning models used in side-channel attacks increases the difficulty of evaluation. If the attack is unsuccessful, the question is if we are dealing with a resistant implementation or a faulty model. We propose an early stopping algorithm that reliably recognizes the model's optimal state during training. The novelty of our solution is an efficient implementation of guessing entropy estimation. Additionally, we formalize two conditions, persistence and patience, for a deep learning model to be optimal. As a result, the model converges with fewer traces.

\end{abstract}
\section{Introduction}

In recent years, deep learning (DL) techniques~\cite{Ian2016grid} have been widely adopted to evaluate the resilience of a cryptographic implementation against side-channel attacks. Several milestones have been reached, and DL became a mainstream side-channel analysis (SCA) evaluation technique. Nevertheless, several aspects still have no satisfactory solution. For instance, preventing the model from underfitting/overfitting is essential because these two phenomena make any DL model perform poorly. In the context of SCA, it represents an important issue because it is uncertain if we are dealing with a resistant cryptographic implementation or a faulty DL model.

Undertraining/overtraining are the leading causes of underfitting/ overfitting\footnote{In the following, we will use only overfitting and overtraining for brevity}. One of the challenges in using DL for SCA is stopping the training process. A suboptimal DL model leads to a suboptimal leakage evaluation, which results in inconclusive evaluation results. Solving the overtraining problem is not trivial, as known DL metrics do not  match the metrics used for SCA. For example, \textit{accuracy} is a metric used to monitor the model state through the training process, and the strategy is to stop the process once the model reaches the desirable accuracy.  However, neither accuracy nor any other known ML metric can be used as a monitor metric for DL model in SCA~\cite{Picek2018TheCO}.

Currently, two approaches exist in solving this problem; Perrin et al. ~\cite{Perin2020LearningWT} use mutual information (MI) as a metric to monitor the training. While effective, its computation cost is a significant drawback. Robissout et al. ~\cite{Damien2020OnlinePerformance} use Guessing Entropy (GE), traditionally a metric used to evaluate the performance of an attack,  as  monitoring metric. As GE is not stable, their strategy is not reliable. Additionally, GE  has a high computation cost, similar to MI~\cite{Perin2020LearningWT}.

We propose an early stopping strategy for DL models used in SCA. Our proposal uses an efficient GE implementation as a metric to monitor the model and stop the training when the model has reached an optimal state. To control the desired model, we define: (i) \textit{persistence} an attribute that monitors the GE convergence in terms false positive outcomes and (ii) \textit{patience} an attribute that controls the confidence 
 in the optimality of the model.

\subsection{Paper contributions are as follows} 
\begin{itemize}
\item We introduce an early stopping mechanism to monitor DL models used in SCA evaluations. Our early stopping strategy reliably recognizes the model's optimal state during training; consequently, we increase the chance to assess the leakage evaluation properly. Our results demonstrate that state-of-the-art models sub-optimally evaluate the leakage traces due to overfitting; our algorithm stops the training process at the model optimal state, resulting in a GE converging with fewer traces. 
\item We designed a new algorithm focused on improving computation time convenient for practical applications to efficiently integrate guessing entropy into the training process.
\item We developed a customized version of a \textit{grid search} technique compatible with our early stopping strategy. This grid search is interrupted when our early strategy declares a DL model optimal (according to the evaluator's expectation), preventing the technique from performing additional and unnecessary training processes. We used our grid search to test the greedy case of our experiments (Sect.~\ref{sec:experimental_result}).
\end{itemize}

\subsection{Paper organization} 
Sect.~\ref{sec:backgound} briefly introduces background on SCA and profiled side-channel attacks. Related works are mentioned in Sect.~\ref{sec:related_works}. Sect.~\ref{sec:datasets} introduces the dataset used in the experiments. Sect.~\ref{sec:sca-es} discusses the main contribution of this paper. Finally, the Sect.~\ref{sec:experimental_result} gives results of our experiments while Sect.~\ref{sec:conclusion_perspectives} concludes the paper.

\subsection{Implementation code} 
The implementation's code repository will be available upon paper acceptance.

\section{Background}
\label{sec:backgound}

\subsection{Side-channel analysis}
Side-channel analysis is a process that comprises three stages; (i) the acquisition of side-channel information, also known as leakage traces or traces (denoted by $\mathbf{t}$), (ii) pre-processing the traces, and (iii) leakage analysis and conducting side-channel attacks. An evaluator (or adversary) performs these three stages to evaluate if the side-channel information leaks sensitive information that might compromise the keys from the target cryptographic implementation. The second stage of pre-processing is not mandatory but can be helpful for the attack. The third stage consists of statistical analysis, leakage assessment, or conducting side-channel analysis using ``classical'' approaches such as Differential Power Analysis (DPA)~\cite{kocher1999dpa} or deep learning based attacks~\cite{maghrebi2016breaking}.

\subsection{Deep learning profiled side-channel attack}
This type of side-channel attack requires a deep learning model designed for classification. Given a model, an iterative training process fits the hyper-parameters by updating them during a stage called back-propagation. We feed the model with batches of profiling traces that we denoted as $\mathcal{P}$. After feed-forwarding the traces through the model, a loss function computes the error between the models' predictions and the ground truth. This stage is repeated through several iterations known as \emph{epochs}. The overall process is called \textit{profiling phase}\footnote{It is called profiling because the evaluator takes a clone of the target device to gather the data used in the training process} of the SCA evaluation.

The \emph{set of attack traces} denoted as $\mathcal{A}$ is used to evaluate the performance of the model. From the set $\mathcal{A}$ a batch of $N_{a}$ attack traces is selected (normally random). The model generates a \emph{prediction vector} $\mathbf{P_r}$ using the batch $N_a$. An \emph{average guessing vector} $\mathbf{g}$, for different sets of $N_a \in \mathcal{A}$ is built as follows: $\mathbf{g} = sort(\mathbb{E}[log(\mathbf{P_r})])$ where $\mathbb{E}$ is the expectation for the different values of $\mathbf{P_r}$. Given $\mathbf{g}$, the GE function is defined as: $\textrm{GE}(\mathbf{g}) = rank_{k^*}(\mathbf{g})$ where $rank_{k*}$ is the rank of the correct key $k*$.

During the training process of an SCA deep learning base model, there is no way to evaluate its performance using e.g. \textit{accuracy} as in other applications of deep learning. However, we can use guessing entropy to resemble what accuracy does. Actually, the guessing entropy evaluates how leaky is the cryptographic implementation on the hand, and it also evaluates the performance of the deep learning model for SCA on the other hand. Given that, we can stop training after recognizing the model met conditions according to a value of the guessing entropy, preventing overtraining.

\section{Related works}
\label{sec:related_works}
Only a few studies have addressed early stopping mechanisms in the context of side-channel analysis. The closest work we identified is from Robissout \textit{et al}.~\cite{Damien2020OnlinePerformance}; they developed an early stopping strategy that computes the rank of a key (using the GE algorithm). Their results suggested that a DL model reaches its optimal state when the rank is close to the lower value of the averaged guessing entropy. Perin \textit{et al}.~\cite{Perin2020LearningWT} proposed an early stopping strategy using mutual information as the monitor metric. Treating a DL model as a Markov chain, they suggested that the amount of secret information remaining (after passing through the layers of the model) points out at its optimal state~\cite{shwartz2017opening}. \textit{The drawbacks of their approach is the computation time of mutual information, which restricts its practical application. Regarding to~\cite{Damien2020OnlinePerformance}, that approach lacks reliability, because the window within their strategy claimed the model reached its optimal state is too short; consequently, it has a high probability to outcome a false positive.}

\section{ASCAD fixed key dataset}
\label{sec:datasets}
ASCAD fixed key\footnote{Publicly available at https://github.com/ANSSI-FR/ASCAD} ($\text{ASCAD}^F$) was introduced in~\cite{Prouff2018StudyOD}. The device used as a clone device is an Atmega8515 8-bit microcontroller, while AES-128 is the cryptographic algorithm protected using masking countermeasure~\cite{joan2002DesignRijndael,Bloemer2004}. Conveniently, the leakage traces are pre-processed to include only the relevant part of the cryptographic algorithm execution, specifically the third masked Sbox in the first round. The dataset's structure allocates leakage traces in two sets; (i) profiling traces ($\mathcal{P}$) comprise $50\,000$ traces, and (ii) attack traces ($\mathcal{A}$) comprise $10\,000$ traces.

\section{Optimized early stopping strategy}
\label{sec:sca-es}
Before discussing the details about our early stopping strategy, we first give some thoughts on the optimization of the GE algorithm. The rest of the section explains our proposed strategy and the use cases in practical applications for our early stopping strategy.

\subsection{Guessing entropy optimization}
Guessing entropy is expensive in terms of computation time and this represents a drawback for integrating it into a training process. We designed an optimized version of its algorithm to overcome the drawback and efficiently integrate the GE metric into the training process. Starting from the algorithm suggested in~\cite{Prouff2018StudyOD}, we reduced the computation time by: (i) removing nested loops and (ii) applying per-block operation using linear algebra. Note that we provide the source code implementation of our strategy, and to save space, we do not include a pseudo-code in the paper.

\begin{figure}[!ht]
\centering
\includegraphics[width=3in]{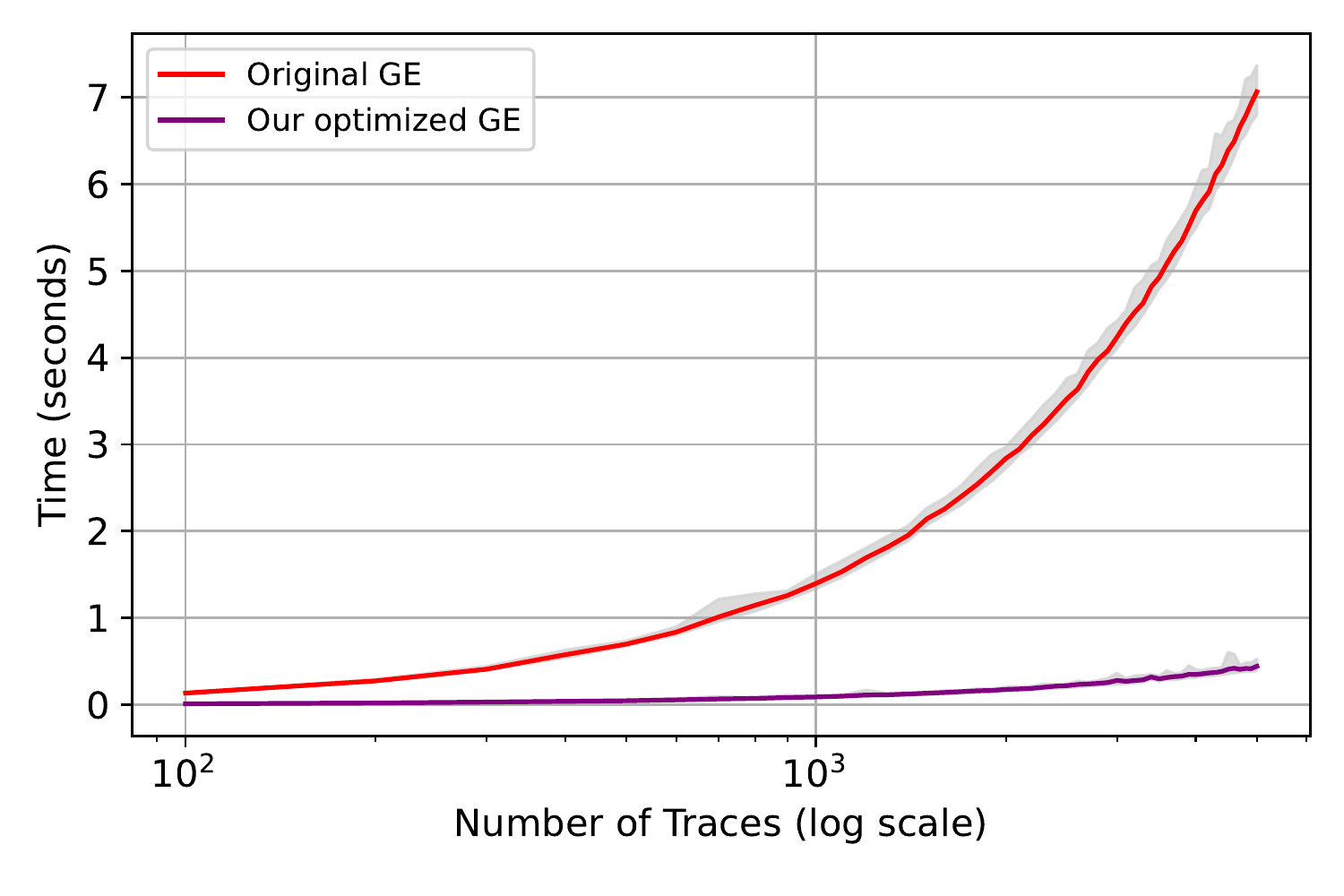}
\caption{Comparison between guessing entropy implementation from~\cite{Prouff2018StudyOD} and our optimized guessing entropy algorithm. Our proposal keeps the computation time below $1$ second for more than $5\,000$ traces}
\label{fig:optimized_and_original_ge}
\end{figure}

By removing nested loops, we pass from an algorithm complexity of order four to one of order three. The Per-block operations organize data into vectors; consequently, we can process this data using linear algebra. In particular, the original algorithm computes the bitwise xor operation per trace required to compare all bytes from the keyspace $\mathcal{K}$ as $k=\{0,\dots |\mathcal{K}|\}$ with the trace label\footnote{For instance, if the cryptographic primitive is the AES S-box, then keyspace $\mathcal{K}=\{0, \cdots 255\}$ and consequently, $\text{S-box}(\{0, 255\} \oplus label\_trace_{i})$} being $|\mathcal{K}|$ the cardinality of the keyspace, and this is done byte per byte. 

In contrast, our algorithm defines a vector of all bytes ($\vec{k} = [0,\cdots, \mathcal{K}]$) and XORes the trace's label as a scalar multiplication substituting the per-value by a per-block operation. Then, the vector feeds the algorithm that computes the key rank ($rank_{k*}$), also using a per-block operation. The resulting throughput has a higher capability than the per-value approach. Fig.~\ref{fig:optimized_and_original_ge} compares the computation time of the implementation from~\cite{Prouff2018StudyOD} and our proposal. We test the algorithm by performing $10$ trials of an experiment that uses $N_{a}=5\,000$ traces, and we measure the computation time in steps of $100$ traces\footnote{Tested it in a Laptop PC Intel(R) Core(TM) i7-9850H CPU\@2.60GHz 2.59 GHz, 16,0 GB RAM, Windows 10 Pro x64 OS}. Clearly, our algorithm outperforms implementation from~\cite{Prouff2018StudyOD} while our proposal keeps the computation time below $1$ second for more than $5\,000$ traces the implementation in~\cite{Prouff2018StudyOD} quickly increases computation time within fewer traces.

Although we achieved a good performance using the same programming language\footnote{Python programming language} as in~\cite{Prouff2018StudyOD}, we consider that it is possible to improve further the performance by using a more efficient programming language. We leave this for future work in optimizing our algorithm.

\subsection{Early stopping algorithm overview}
\begin{figure}[!ht]
\centering
\includegraphics[width=3in]{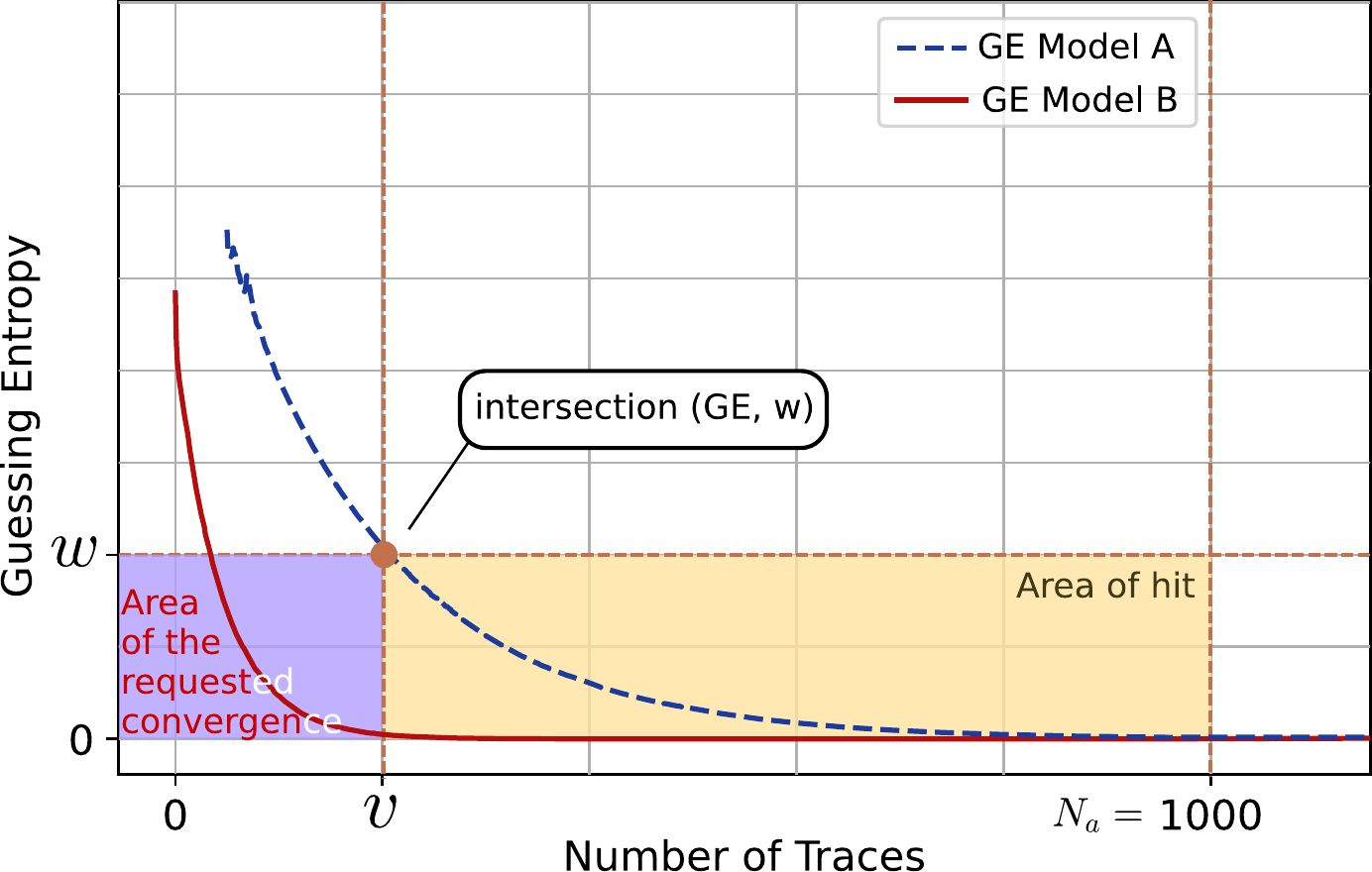}
\caption{Persistence monitors GE through the area of hit. If the GE persists within the area up to $N_a$ the strategy indicates that the model hit the desirable performance.}
\label{fig:area_of_hits}
\end{figure}

A well-trained deep learning model for SCA evaluation would result in a guessing entropy that converges to zero. If we represent this convergence as the limit when the GE curve approaches to a right side-opened interval of traces, defined by a lower bound $v$ up to $N_{a}$, \textit{i.e.},
\begin{math}
  \lim_{\mathbf{t} \rightarrow [v, N_{a}]}\text{GE}[\mathbf{t}]=0
\end{math}; then, we can find a GE value $w$ where the limit is met. More precisely, the limit suggests that after a specific number of traces $v$, the guessing entropy reaches its maximum of convergence (being zero, in the best case scenario). However, the limit also suggests that the convergence \textit{persists} up to a maximum of traces and beyond. The \textbf{Fig.~\ref{fig:area_of_hits}} illustrates this with a number of traces $N_{a}=1\,000$.

Furthermore, by setting $w$, we can find $v$ by projecting, onto the number of traces axis, the intersection between the GE curve and the line that passes through $w$ (see \textbf{Fig.~\ref{fig:area_of_hits}} GE Model A). With these three parameters $(w, v, N_{a})$, we define the \textit{area of hit} as the region where the GE curve should persist up to $N_{a}$ number of traces. From this analysis, we derive the first criterion of our early stopping strategy;

\begin{definition}{\textbf{Persistence}:}\label{def:persistence}
the function denoted by
$\mathbf{P_e}(w, N_{a}) = v\,\text{such that}$ $v,w,\text{and}\,N_{a} \in \mathbb{Z}$ defines the area of hit where the GE curve should persist toward meeting the stop conditions.
\end{definition}

When the conditions fixed by the persistence are met, our early stopping strategy declares that the model has accomplished a GE value according to the evaluator's expectations. A parameter $w$ makes sense for those cases when the number of attack traces $N_{a}$ is limited. For instance, let us take $\text{ASCAD}^F$ maximum number of attack traces ($10\,000$). Then, a classifier aimed to evaluate those leakage traces might generate a GE not converging towards zero, but some other positive value, before $10\,000$ traces. Consequently, the early stopping strategy would never trigger the evaluator's conditions unless he/she changes the parameter $w$ of $\mathbf{P_e}$. 

Now, our early stopping strategy should monitor the stability of the GE convergence to avoid any false positives. Stability in this context refers to the times the persistence frames the GE curve, leading us to the concept of patience.

\begin{definition}{\textbf{Patience} ($P_a$)}\label{def:patience}
is a value that establishes the number of consecutive epochs for the guessing entropy to stay in the area of hit.
\end{definition}
In other words, if the model does hit the persistence $P_a$ times, then the early stopping strategy does stop training since the stability criterion was met. Although patience is a common parameter in most early stopping algorithms, the patience from our early stopping strategy is subject to the number of attacks performed in each epoch. Our experiments show that performing several attacks brings the model under more epochs, because meeting the patience criterion becomes more difficult. We will come to this later in this paper.

\subsubsection{The soft and greedy use cases} Notice that Fig.~\ref{fig:area_of_hits} has two GE curves from Model A and B, intending to illustrate two possible use cases of the persistence. The use case from GE Model A is when the evaluator sets the parameters $w$ and $N_{a}$; wherever the GE Model A intersects with $w$-line, we get the value of $v$ and the area of hit. This case is called \textit{soft case} because $v$ takes different values through training epochs; yet, they count as persistence and patience hit.

However, let us suppose we have to outperform GE Model A result, meaning we should find a deep learning model whose GE converges with fewer traces than Model A (like GE Model B). In that case, our proposed strategy allows the evaluator to set parameter $v$ in advance. When $v$ is previously set, its value does not change through training epochs; moreover, the area of hit merges with the \textit{area of requested convergence}. Given that, the guessing entropy should now touch both areas to claim a persistence and patience hit. It is called \textit{greedy case} because it represents a more challenging goal to achieve.

\subsubsection{Finding optimality with grid search} A good approach is to combine the greedy use case with an ``optimal model searching algorithm'' such as \textit{grid search}\footnote{As well as \textit{random search}, or \textit{bayes search}~\cite{Bergstra2012RandomSearch}}~\cite{Ian2016grid}. Grid search is a well-known algorithm used in machine learning that might derive an optimal model from a set of hyper-parameters. As shown by~\cite{paguada2020controlling}, grid search has the limitation that there are no implemented metrics toward seeking optimal models for SCA. Consequently, the common practice is to let the grid search pass through the whole set of hyper-parameters and  then evaluate the performance of all possible models leading to an inefficient practice. Combining our strategy and grid search, one can efficiently stop the search, avoiding further computation after finding the best model. We will discuss this in more detail in the experiments section (Sect.~\ref{sec:experimental_result}).

\subsubsection{Persistence modes} We feature the persistence in two modes; (i) \textit{full} and (ii) \textit{binary} persistence. Full persistence acts exactly how Definition~\ref{def:persistence} conceptualizes the persistence, \textit{i.e.}, to claim a hit, the GE curve should not go outside the area of hit at any number of traces. In contrast, binary persistence allows defining a percentage of traces the GE curve should keep in the area of hit\footnote{The number of traces not necessarily being continuous}. For instance, a value of $0.95$ in binary persistence mode means that $5\%$ of traces are allowed outside the area of hit. Binary persistence helps in situations when the GE takes a wrong key as the correct key; as a result, it starts diverging~\cite{wu2020attack}. So, to bypass this error, we included the binary persistence, and we let the full persistence as an option to support the ``traditional'' approach where GE should always be descending.

\section{Experimental results}
\label{sec:experimental_result}
This section discusses our experimental results. We split the experiments into the two use cases issued in the previous section. In the soft case, we compare our strategy with the early stopping strategy defined in~\cite{Damien2020OnlinePerformance} (we called it key-rank strategy).

\subsection{Case 1: Soft case}
This experiment uses the model defined in~\cite{Zaid2019} trained with $\text{ASCAD}^F$, according to the study, the model's GE converges to zero after 191 traces using 50 epochs, a batch size of 50, and using \textit{One cycle policy} to control the learning rate. We called this model Model\_v1.

\begin{table}[!ht]
\centering
\begin{tabular}{lr}
\toprule
Layer type&Details\\
\midrule
Conv & \# kernels 4, kernel size 1, SeLU\\
Pooling & Average, kernel size 2\\
Batch normalization &\\
\midrule
Fully-connected & \# units 10, SeLU \\
Fully-connected & \# units 10, SeLU \\
Fully-connected & \# units 256, Softmax \\
\bottomrule
\end{tabular}
\caption{Deep learning architecture summary for Model\_v1, each table row represents a layer of the model.}
\label{tab:model_v1}
\end{table}

\textbf{Table~\ref{tab:model_v1}} summarizes the Model\_v1 architecture. For training it, we used $|\mathcal{P}| = 45\,000$. Values of the learning rate, epoch, and batch size as in the original work. Our early stopping strategy requires attack traces, so we took $|\mathcal{A}|=10\,000$; then, we set parameters\footnote{$w=0$ means that we expect the model to achieve a GE with the commonly requested level of convergence} $N_a=5\,000$, $w=0$ and $persistence\,mode=full$. Notice that for the sake of completeness, we let the training finishes.

\begin{figure}[!ht]
\centering
\includegraphics[width=3in]{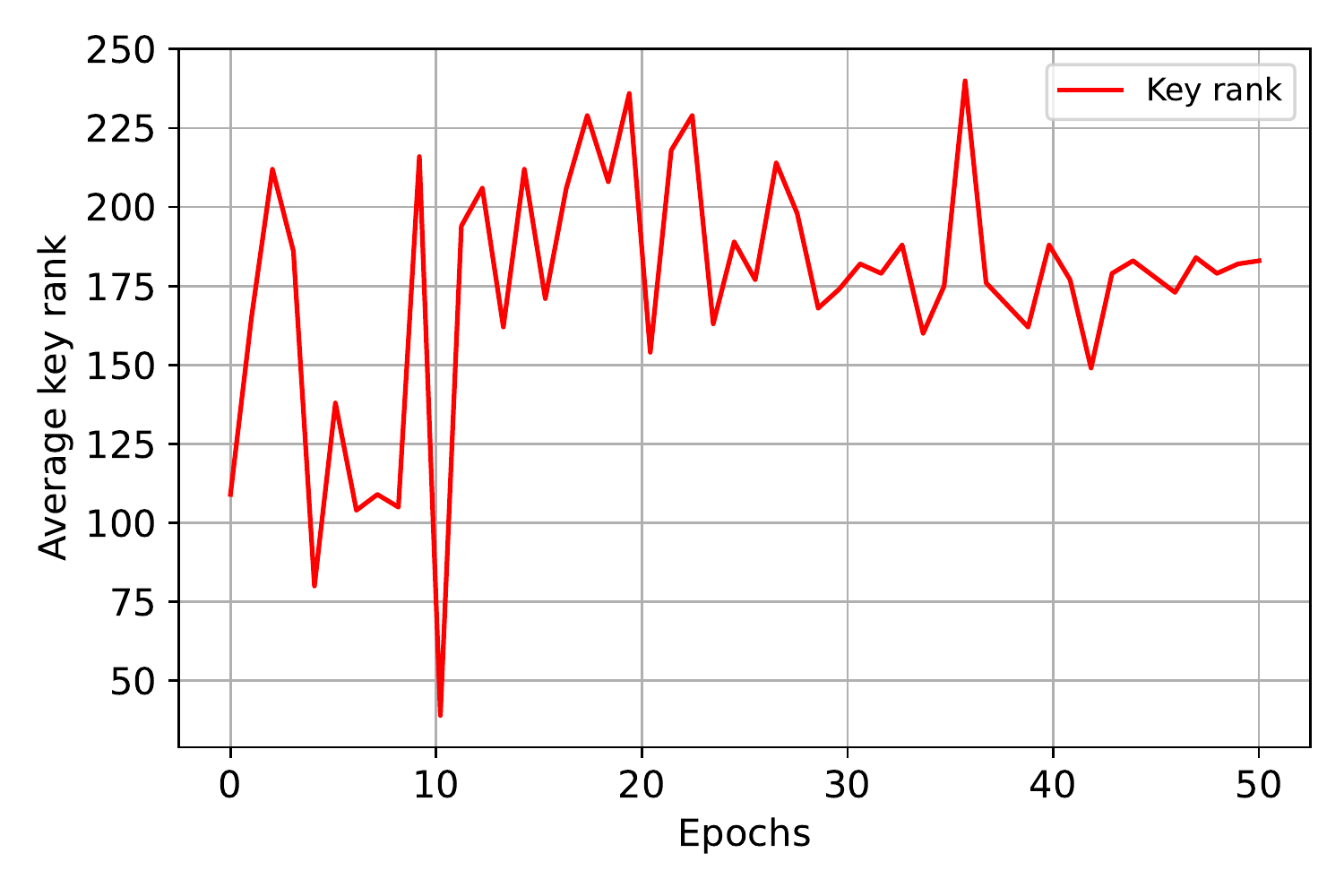}
\caption{the Key-rank strategy suggests that the model reached its optimal state at epoch 10.}
\label{fig:average_key_rank}
\end{figure}

\begin{figure}[!ht]
\centering
\includegraphics[width=3in]{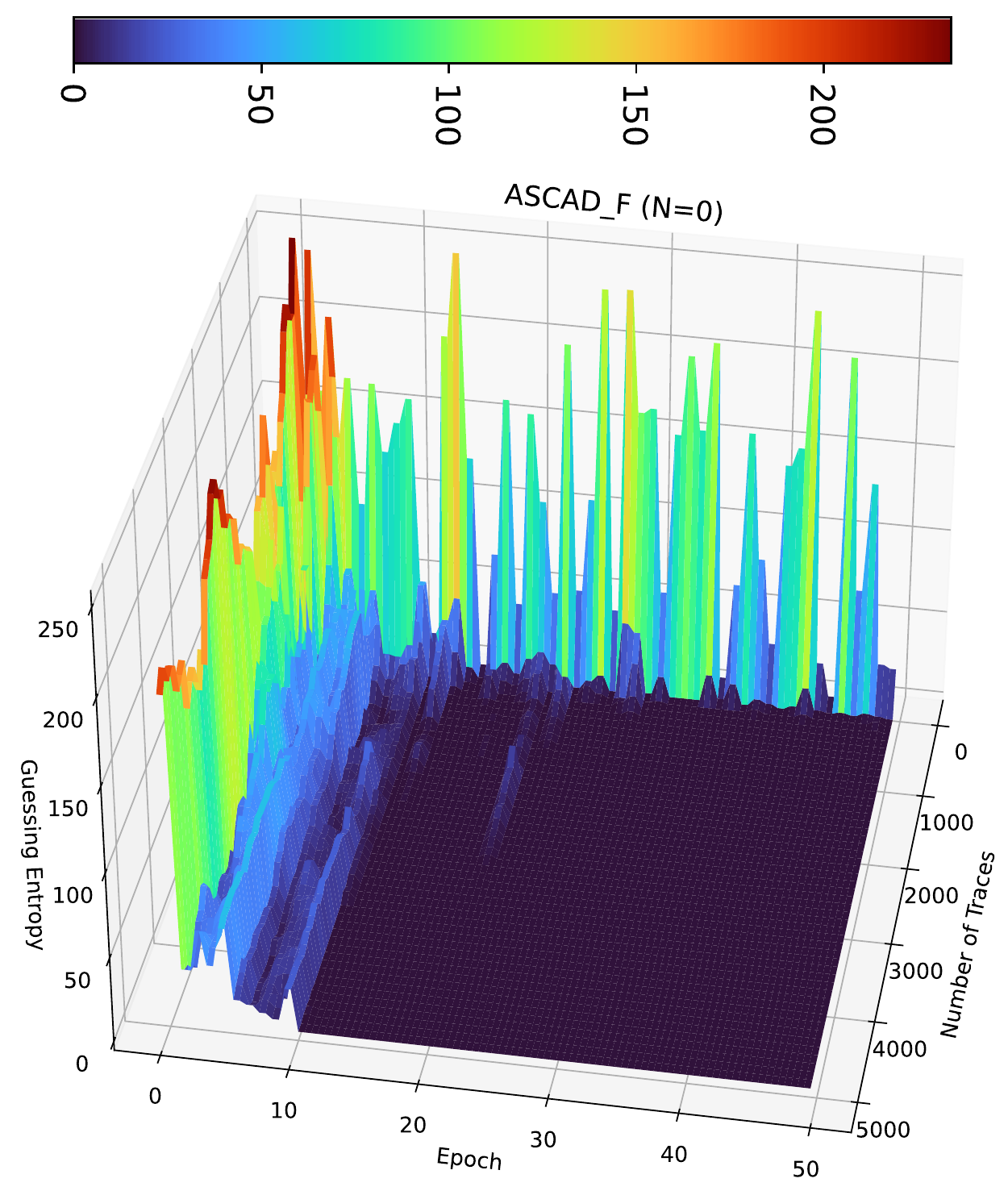}
\caption{The area over the surface that resembles a valley starting from epoch 10 suggests the optimal state of the model.}
\label{fig:ge_surface_epoch}
\end{figure}

\textbf{Fig.~\ref{fig:average_key_rank}} depicts how the key-rank strategy suggests the model has reached its optimal state, while \textbf{Fig.~\ref{fig:ge_surface_epoch}} depicts our strategy results. Notice that the key-rank strategy has the lowest peak after 10 epochs, exactly where the plot in \textbf{Fig.~\ref{fig:ge_surface_epoch}} starts with a valley representing the lower values of the GE.

However, two observations are essential to notice at this point. First, by letting the training continue, we observed that a few epochs ahead (around epoch 13), the model performed better than in epoch 10. Moreover, it kept this convergence up to epoch 20 while key-rank strategy immediately suggests that the model starts moving from this optimal state to a likely overfitting state. This latter suggestion leads us to our second observation: according to the way the evaluator sets the patience of our proposed strategy, a single hit like in key-rank strategy would not stop the training. Actually, by using the key-rank strategy, there is no way to monitor the stability of the convergence leading to uncertain outcomes.

The window of ``time'' that the key-rank strategy has is too short, implying that we cannot ensure having reached the optimal state of the model. For example, let us take the second lower peak at epoch 4 and suppose we decide to stop the training since the following peak has a higher value. However, around none of those points, the model had reached its optimal state. Contrary, setting $P_a=3$ ensures that up to 13 epochs, it will count 3 hits (starting from 10) and stop training since the model met the stability conditions. This experiment demonstrated that our strategy concludes more intuitively and efficiently when the model has reached its optimal state.

\subsection{Case 2: Greedy case}
\begin{figure}[!ht]
\centering
\includegraphics[width=2.8in]{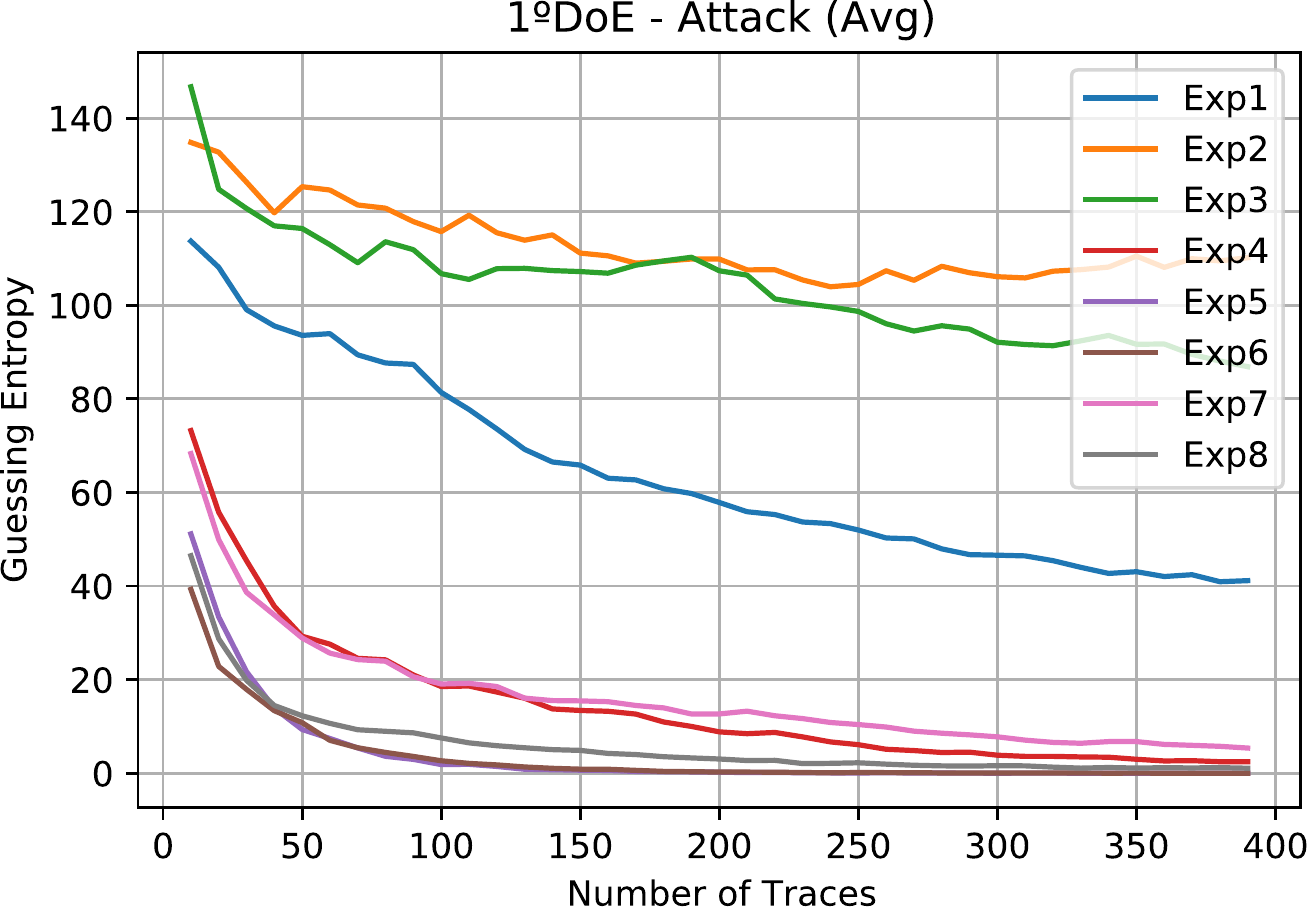}
\caption{According~\cite{paguada2020controlling} in experiments 5 and 6 using 25 epochs GE converges earlier, while in experiments 7 and 8 using 50 epochs GE got higher values. We thank to the authors of~\cite{paguada2020controlling} for allowing us to use Fig.~4 - first iteration of DoE.}
\label{fig:DOE1_GE_PLOT}
\end{figure}

Next, we use the results presented in~\cite{paguada2020controlling} where the authors applied a customized version of Six-sigma methodology~\cite{rioja2020uncertainty,cheng2012} over a standard grid search algorithm toward finding a model whose GE converges earlier than Model\_v1. Their goal was similar to ours since we are trying to optimize the grid search by stopping it when we find a ``better model''. Their result shows that by adding a fully-connected layer (of 10 units), the better model (called Model\_v2) converges with fewer traces, \textit{i.e.} around 150 instead of 191 traces (see \textbf{Fig.~\ref{fig:DOE1_GE_PLOT}}).

Nevertheless, our experiment shows that it is only necessary to train Model\_v1 using the correct number of epochs to outperform its previous result. For this experiment in particular, we define a set of four hyper-parameters; each one takes two possible values, as shown in \textbf{Table~\ref{tab:parameters_space}}. As a result, we build the so-called \textit{hyper-parameters space} of the grid search. Given that each hyper-parameter can take one of two values, we have 16 possible training processes.

\begin{table}[!ht]
\centering
\begin{tabular}{lr}
\toprule
Variable/hyper-parameter&Values\\
\midrule
Architecture & \{Model\_v1, Model\_v2\}\\
Batch size & \{50, 100\}\\
Epochs & \{50, 100\}\\
Optimizer & \{\textit{RMSprop}, \textit{Adam}\} \\
\bottomrule
\end{tabular}
\caption{Variables/Hyper-parameters and values define the hyper-parameters space of the grid search.}
\label{tab:parameters_space}
\end{table}

Our aim is that the early stopping strategy stops the grid search, when any of those training processes derive a model that meets the stop conditions. Clearly, the soft use case would also do in this scenario, but only if the goal is to come across a model whose guessing entropy converges at any number of traces. However, in some circumstances, like looking for a model that outperforms previous ones, we must define a specific minimum of traces, \textit{i.e.} we should set a value for $v$ in advance. In such circumstances, a standard grid search algorithm will go over all 16 training processes, helpful if the evaluator is looking to test all of them; otherwise, it represents a costly process in terms of time. In contrast, our grid search version\footnote{Code available at the same repository} stops searching after a single model achieves the required performance.

\begin{figure}[!ht]
\centering
\includegraphics[width=3in]{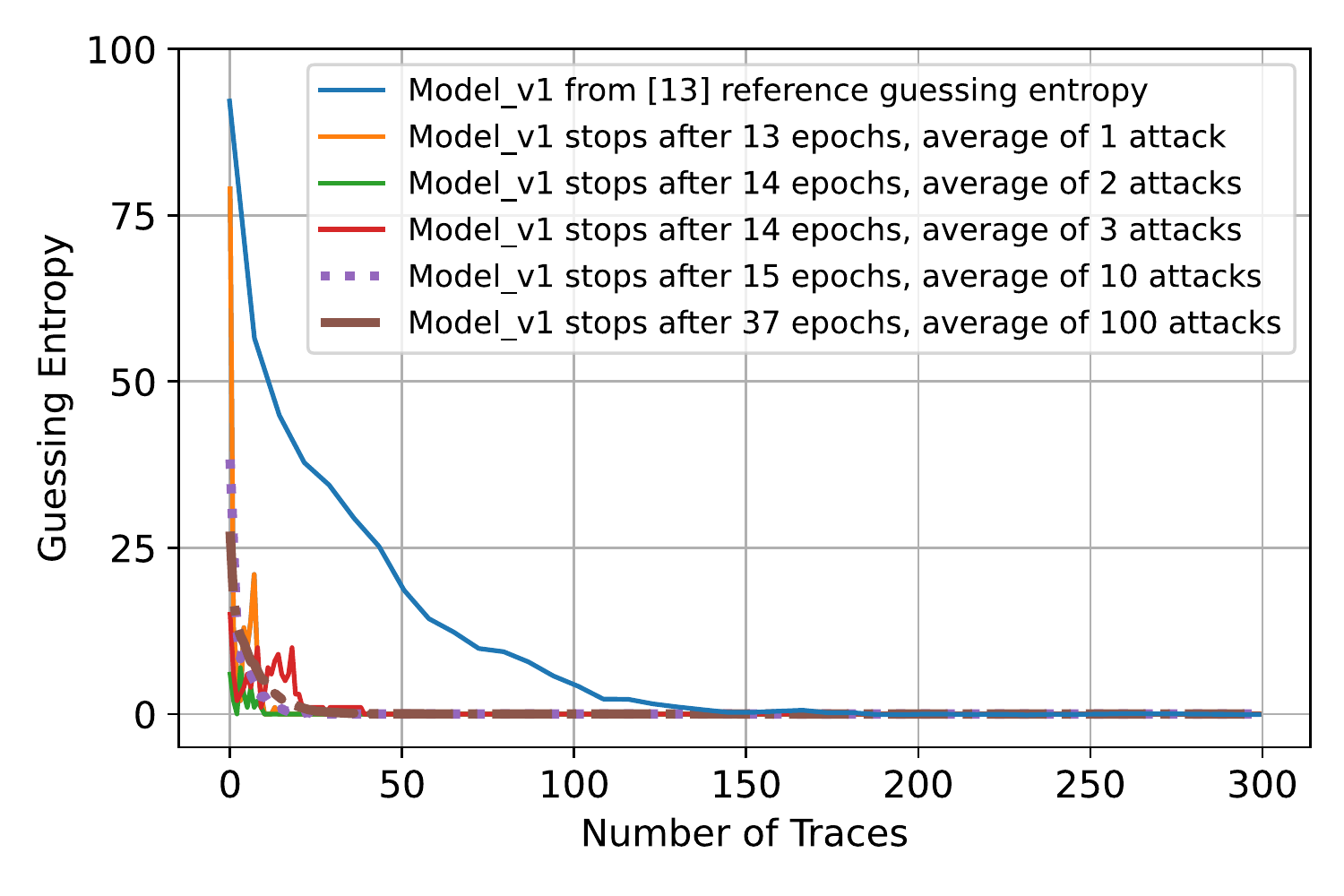}
\caption{Guessing entropy of five training sessions with different number of attacks.}
\label{fig:grid_search_sca_es}
\end{figure}

For our experiment, we fix $v = 100$ as the number of traces where the guessing entropy should already be less than $w$ (touching the area of the requested convergence) and $persistence\,mode=binary$ with a value of $0.95$. Fig.~\ref{fig:grid_search_sca_es} depicts the result. Note that we repeat five times the grid search by letting our strategy perform different numbers of attacks to average the GE curve. Indeed, an averaged guessing entropy encourages the stability criterion. However, the number of epochs increases according to the number of attacks because of the variance. Consequently, it could quickly end in overfitting the model due to the additional epochs. This experiment shows that as long as the number of attacks by epoch guarantees the stability of the convergence, a few attacks are enough~\cite{Wu2021OntheEvaluation}.

In this particular result, all five repetitions are from the first training process (first combination of hyper-parameters), and all of them outperform the GE reference from~\cite{Zaid2019} with less than 40 traces in less than 50 epochs. Since the ``best'' model (Model\_v1 architecture) was in the first training process, we stopped the grid search after 1 training process out of 16 possible ones (best case scenario). Results from~\cite{paguada2020controlling} support this experiment's conclusions; notice that according to their results, those models trained using 25 epochs performed better than models trained using 50 (see \textbf{Fig.~\ref{fig:DOE1_GE_PLOT}}). Furthermore, our results manifest that we also outperformed the result from~\cite{paguada2020controlling} in both searching time and GE convergence. Clearly, the overfitting was the problem that limited the model's performance in both works~\cite{paguada2020controlling} and~\cite{Zaid2019}.

\section{Conclusions and future works}
\label{sec:conclusion_perspectives}

This paper introduced an optimized early stopping strategy for deep learning models used in side-channel attacks. Our proposal defined patience and persistence as criteria for monitoring the guessing entropy from two different axes; (i) its stability through the training epochs and (ii) through the number of attack traces. Our proposal reliably recognizes when the deep learning model reaches its optimal state by keeping track of the guessing entropy from these two different perspectives. It prevents the model from overfitting, reducing the uncertainty of getting a faulty model.

Our proposal relies on the guessing entropy. Researches to this date evidence that guessing entropy is a feasible side-channel analysis metric to evaluate the performance of an attack. However, the metric might exhibit false outcomes during a side-channel evaluation; on top of that, it might be considered that its time complexity is too heavy to be computed during training. Our early stopping strategy assists the evaluator in walking around those issues. To overcome computation time, we developed an optimized version of the guessing entropy algorithm. Concerning false outcomes, our proposed strategy can monitor the guessing entropy in two different modes of persistence, allowing the evaluator to design stopping conditions that best suit the case.

This work clears the way for further studies to improve or develop new early stopping strategies based on the guessing entropy. In future works, we plan to incur other optimization techniques to reduce the computation time even more. Finally, we consider that our early stopping strategy algorithm can be the base for a score function for more efficient hyper-parameter searching algorithms.
%
%
%
\bibliographystyle{splncs04}
\bibliography{biblio}
\end{document}